\documentclass[12pt,a4paper]{article}

\usepackage{ifthen} 
\newboolean{pdflatex}
\setboolean{pdflatex}{true} 

\newboolean{articletitles}
\setboolean{articletitles}{true} 

\newboolean{uprightparticles}
\setboolean{uprightparticles}{true} 

\newboolean{inbibliography}
\setboolean{inbibliography}{false} 

\usepackage{rotating} 
\usepackage{graphpap}
\usepackage{float}
\usepackage{placeins}
\usepackage{afterpage}

\usepackage{microtype}
\usepackage{lineno}  
\usepackage{xspace} 

\usepackage{graphicx}  
\usepackage{color}
\usepackage{colortbl}
\graphicspath{{./figs/}} 

\usepackage{amsmath} 
\usepackage{amssymb}
\usepackage{amsfonts}
\usepackage{upgreek} 

\newcommand*\patchAmsMathEnvironmentForLineno[1]{%
\expandafter\let\csname old#1\expandafter\endcsname\csname #1\endcsname
\expandafter\let\csname oldend#1\expandafter\endcsname\csname
end#1\endcsname
 \renewenvironment{#1}%
   {\linenomath\csname old#1\endcsname}%
   {\csname oldend#1\endcsname\endlinenomath}%
}
\newcommand*\patchBothAmsMathEnvironmentsForLineno[1]{%
  \patchAmsMathEnvironmentForLineno{#1}%
  \patchAmsMathEnvironmentForLineno{#1*}%
}
\AtBeginDocument{%
\patchBothAmsMathEnvironmentsForLineno{equation}%
\patchBothAmsMathEnvironmentsForLineno{align}%
\patchBothAmsMathEnvironmentsForLineno{flalign}%
\patchBothAmsMathEnvironmentsForLineno{alignat}%
\patchBothAmsMathEnvironmentsForLineno{gather}%
\patchBothAmsMathEnvironmentsForLineno{multline}%
}

\usepackage{hyperref}    
\usepackage[all]{hypcap} 

\def\lhcb {\mbox{LHCb}\xspace}

\ifthenelse{\boolean{uprightparticles}}%
{

 \def\Pvarepsilon {\ensuremath{\upvarepsilon}\xspace}

 \def\Pmu         {\ensuremath{\upmu}\xspace}

 \def\Ppi         {\ensuremath{\uppi}\xspace}

 \def\Ptau        {\ensuremath{\uptau}\xspace}                 
                  
 \def\Pphi        {\ensuremath{\upphi}\xspace}

 \def\Ppsi        {\ensuremath{\uppsi}\xspace}                 
                  
 \def\Psigma      {\ensuremath{\upsigma}\xspace}                 

 \def\PDelta      {\ensuremath{\Delta}\xspace}                 
 \def\PXi      {\ensuremath{\Xi}\xspace}                 
 \def\PLambda      {\ensuremath{\Lambda}\xspace}                 
 \def\PSigma      {\ensuremath{\Sigma}\xspace}                 
 \def\POmega      {\ensuremath{\Omega}\xspace}                 
 \def\PUpsilon      {\ensuremath{\Upsilon}\xspace}                 
 
 \def\PB      {\ensuremath{\mathrm{B}}\xspace}                 
                  
 \def\PD      {\ensuremath{\mathrm{D}}\xspace}

 \def\PJ      {\ensuremath{\mathrm{J}}\xspace}                 
 \def\PK      {\ensuremath{\mathrm{K}}\xspace}

 \def\PS      {\ensuremath{\mathrm{S}}\xspace}

 \def\PW      {\ensuremath{\mathrm{W}}\xspace}

 \def\Pb      {\ensuremath{\mathrm{b}}\xspace}                 
 \def\Pc      {\ensuremath{\mathrm{c}}\xspace}

 \def\Pi      {\ensuremath{\mathrm{i}}\xspace}

 \def\Pp      {\ensuremath{\mathrm{p}}\xspace}

 \def\Ps      {\ensuremath{\mathrm{s}}\xspace}

}
{

 \def\Pvarepsilon {\ensuremath{\varepsilon}\xspace}

 \def\Pmu         {\ensuremath{\mu}\xspace}

 \def\Ppi         {\ensuremath{\pi}\xspace}

 \def\Ptau        {\ensuremath{\tau}\xspace}                 
                  
 \def\Pphi        {\ensuremath{\phi}\xspace}

 \def\Ppsi        {\ensuremath{\psi}\xspace}                 
                  
 \mathchardef\PDelta="7101
 \mathchardef\PXi="7104
 \mathchardef\PLambda="7103
 \mathchardef\PSigma="7106
 \mathchardef\POmega="710A
 \mathchardef\PUpsilon="7107
                  
 \def\PB      {\ensuremath{B}\xspace}                 
                  
 \def\PD      {\ensuremath{D}\xspace}

 \def\PJ      {\ensuremath{J}\xspace}                 
 \def\PK      {\ensuremath{K}\xspace}

 \def\PS      {\ensuremath{S}\xspace}

 \def\PW      {\ensuremath{W}\xspace}

 \def\Pb      {\ensuremath{b}\xspace}                 
 \def\Pc      {\ensuremath{c}\xspace}

 \def\Pi      {\ensuremath{i}\xspace}

 \def\Pp      {\ensuremath{p}\xspace}

 \def\Ps      {\ensuremath{s}\xspace}

}

\def\mumu       {\ensuremath{\Pmu^+\Pmu^-}\xspace}

\def\W      {\ensuremath{\PW}\xspace}
\def\Wp     {\ensuremath{\PW^+}\xspace}

\def\squark    {\ensuremath{\Ps}\xspace}

\def\cquark    {\ensuremath{\Pc}\xspace}

\def\bquark    {\ensuremath{\Pb}\xspace}

\def\pion  {\ensuremath{\Ppi}\xspace}

\def\pip   {\ensuremath{\pion^+}\xspace}
\def\pim   {\ensuremath{\pion^-}\xspace}

\def\kaon  {\ensuremath{\PK}\xspace}
  \def\Kbar  {\kern 0.2em\overline{\kern -0.2em \PK}{}\xspace}

\def\Kp    {\ensuremath{\kaon^+}\xspace}
\def\Km    {\ensuremath{\kaon^-}\xspace}

  \def\Dbar    {\kern 0.2em\overline{\kern -0.2em \PD}{}\xspace}
\def\D       {\ensuremath{\PD}\xspace}

\def\Ds      {\ensuremath{\D^+_\squark}\xspace}
\def\Dsp     {\ensuremath{\D^+_\squark}\xspace}

\def\B       {\ensuremath{\PB}\xspace}
\def\Bbar    {\ensuremath{\kern 0.18em\overline{\kern -0.18em \PB}{}}\xspace}

\def\Bu      {\ensuremath{\B^+}\xspace}

\def\Bp      {\ensuremath{\Bu}\xspace}

\def\Bd      {\ensuremath{\B^0}\xspace}
\def\Bs      {\ensuremath{\B^0_\squark}\xspace}

\def\Bc      {\ensuremath{\B_\cquark^+}\xspace}

\def\jpsi     {\ensuremath{{\PJ\mskip -3mu/\mskip -2mu\Ppsi\mskip 2mu}}\xspace}

  \def\Y#1S{\ensuremath{\PUpsilon{(#1S)}}\xspace}

\def\proton      {\ensuremath{\Pp}\xspace}

\def\Lbar {\ensuremath{\kern 0.1em\overline{\kern -0.1em\PLambda}}\xspace}

\def\BF         {{\ensuremath{\cal B}\xspace}}

\def\BR         {\BF}

\def\to                 {\ensuremath{\rightarrow}\xspace}

\def\AT#1     {\ensuremath{A_{\mathrm{T}}^{#1}}\xspace}           

\def\C#1      {\ensuremath{\mathcal{C}_{#1}}\xspace}                       
\def\Cp#1     {\ensuremath{\mathcal{C}_{#1}^{'}}\xspace}                    
\def\Ceff#1   {\ensuremath{\mathcal{C}_{#1}^{\mathrm{(eff)}}}\xspace}        
\def\Cpeff#1  {\ensuremath{\mathcal{C}_{#1}^{'\mathrm{(eff)}}}\xspace}       
\def\Ope#1    {\ensuremath{\mathcal{O}_{#1}}\xspace}                       
\def\Opep#1   {\ensuremath{\mathcal{O}_{#1}^{'}}\xspace}                    

\newcommand{\tev}{\ifthenelse{\boolean{inbibliography}}{\ensuremath{~T\kern -0.05em eV}\xspace}{\ensuremath{\mathrm{\,Te\kern -0.1em V}}\xspace}}
\newcommand{\gev}{\ensuremath{\mathrm{\,Ge\kern -0.1em V}}\xspace}
\newcommand{\mev}{\ensuremath{\mathrm{\,Me\kern -0.1em V}}\xspace}
\newcommand{\kev}{\ensuremath{\mathrm{\,ke\kern -0.1em V}}\xspace}
\newcommand{\ev}{\ensuremath{\mathrm{\,e\kern -0.1em V}}\xspace}
\newcommand{\gevc}{\ensuremath{{\mathrm{\,Ge\kern -0.1em V\!/}c}}\xspace}
\newcommand{\mevc}{\ensuremath{{\mathrm{\,Me\kern -0.1em V\!/}c}}\xspace}
\newcommand{\gevcc}{\ensuremath{{\mathrm{\,Ge\kern -0.1em V\!/}c^2}}\xspace}
\newcommand{\gevgevcccc}{\ensuremath{{\mathrm{\,Ge\kern -0.1em V^2\!/}c^4}}\xspace}
\newcommand{\mevcc}{\ensuremath{{\mathrm{\,Me\kern -0.1em V\!/}c^2}}\xspace}

\def\mm   {\ensuremath{\rm \,mm}\xspace}

\def\mum  {\ensuremath{\,\upmu\rm m}\xspace}

\def\invfb   {\ensuremath{\mbox{\,fb}^{-1}}\xspace}

\newcommand{\chisq}{\ensuremath{\chi^2}\xspace}

\newcommand{\chisqvtx}{\ensuremath{\chi^2_{\rm vtx}}\xspace}

\def\gsim{{~\raise.15em\hbox{$>$}\kern-.85em
          \lower.35em\hbox{$\sim$}~}\xspace}
\def\lsim{{~\raise.15em\hbox{$<$}\kern-.85em
          \lower.35em\hbox{$\sim$}~}\xspace}

\def\sPlot{\mbox{\em sPlot}}

\def\pt         {\mbox{$p_{\rm T}$}\xspace}

\def\bcvegpy    {\mbox{\textsc{Bcvegpy}}\xspace}

\def\evtgen     {\mbox{\textsc{EvtGen}}\xspace}

\def\geant      {\mbox{\textsc{Geant4}}\xspace}

\def\photos     {\mbox{\textsc{Photos}}\xspace}

\def\pythia     {\mbox{\textsc{Pythia}}\xspace}

\def\tell1  {TELL1\xspace}
\def\ukl1   {UKL1\xspace}

\usepackage{cite} 
\usepackage{mciteplus}

\usepackage{longtable} 

\begin{document}

\renewcommand{\thefootnote}{\fnsymbol{footnote}}
\setcounter{footnote}{1}


\begin{titlepage}
\pagenumbering{roman}

\vspace*{-1.5cm}
\centerline{\large EUROPEAN ORGANIZATION FOR NUCLEAR RESEARCH (CERN)}
\vspace*{1.5cm}
\hspace*{-0.5cm}
\begin{tabular*}{\linewidth}{lc@{\extracolsep{\fill}}r}
\ifthenelse{\boolean{pdflatex}}
{\vspace*{-2.7cm}\mbox{\!\!\!\includegraphics[width=.14\textwidth]{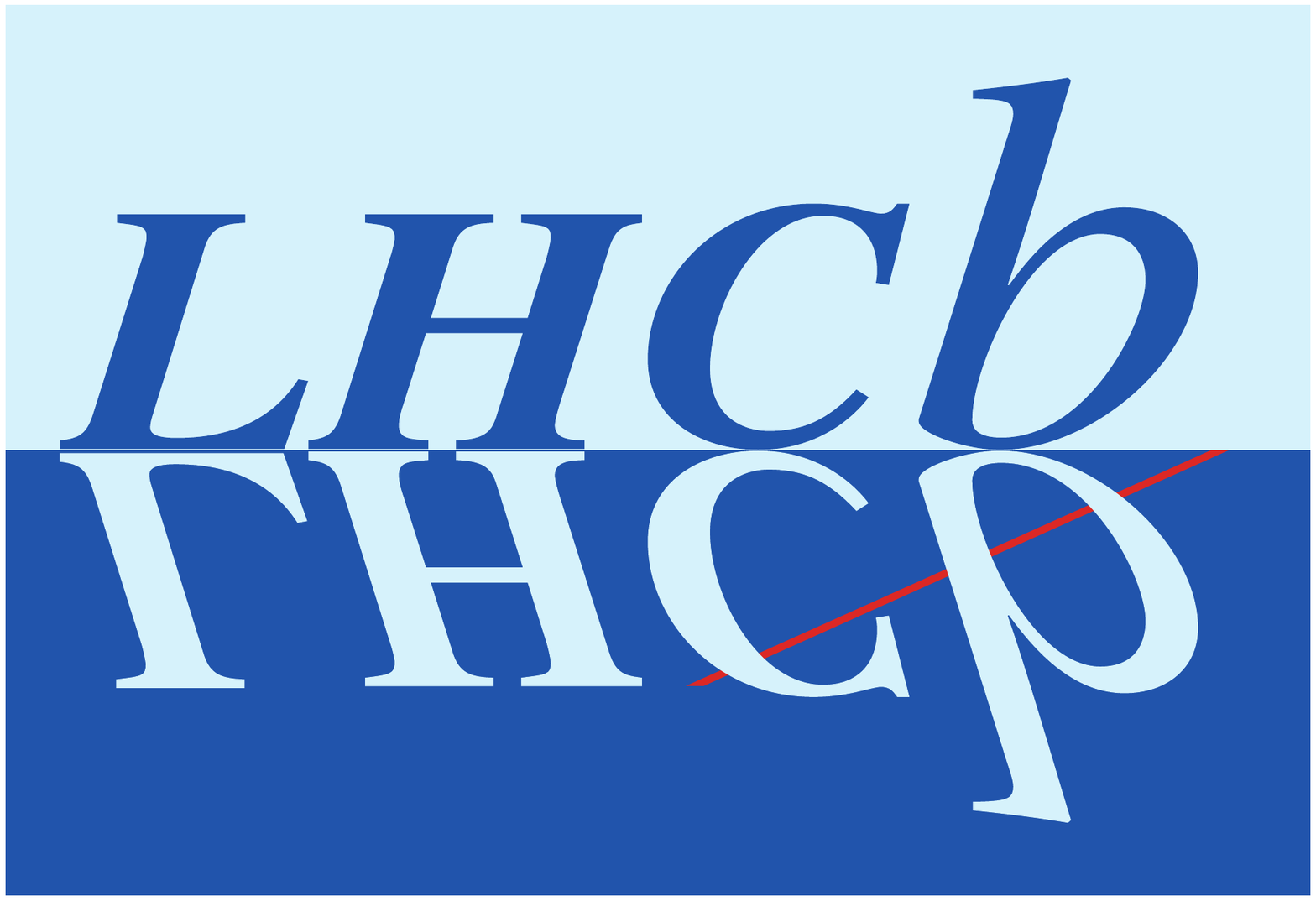}} & &}%
{\vspace*{-1.2cm}\mbox{\!\!\!\includegraphics[width=.12\textwidth]{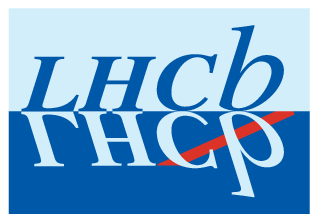}} & &}%
\\
   & & CERN-PH-EP-2013-165 \\  
   & & LHCb-PAPER-2013-047 \\  
   & & September~3, 2013   \\ 
\end{tabular*}

\vspace*{1.5cm}

{\bf\boldmath\huge
\begin{center}
 Observation of the decay $\Bc\to\jpsi\Kp\Km\pip$
\end{center}
}

\vspace*{1.0cm}

\begin{center}
The LHCb collaboration\footnote{Authors are listed on the following pages.}
\end{center}

\vspace{\fill}

\begin{abstract}
  \noindent
The decay $\Bc\to\jpsi\Kp\Km\pip$~is observed for the~first time,
using proton-proton collisions collected with the LHCb~detector
corresponding to an~integrated luminosity of~3\invfb.
A signal yield of~$78\pm14$~decays 
is reported with a~significance of~6.2~standard 
deviations.
The~ratio of the~branching fraction 
of~\mbox{$\Bc\to\jpsi\Kp\Km\pip$}~decays 
to that of~\mbox{$\Bc\to\jpsi\pip$}~decays 
is measured to be
\mbox{$0.53\pm 0.10\pm0.05$},
where the~first uncertainty is statistical and the~second is systematic.
\end{abstract}

\vspace*{1.0cm}

\begin{center}
  Submitted to JHEP 
\end{center}

\vspace{\fill}

{\footnotesize 
\centerline{\copyright~CERN on behalf of the \lhcb collaboration, 
   license \href{http://creativecommons.org/licenses/by/3.0/}{CC-BY-3.0}.}}
\vspace*{2mm}

\end{titlepage}


\newpage
\setcounter{page}{2}
\mbox{~}
\newpage

\centerline{\large\bf LHCb collaboration}
\begin{flushleft}
\small
R.~Aaij$^{40}$, 
B.~Adeva$^{36}$, 
M.~Adinolfi$^{45}$, 
C.~Adrover$^{6}$, 
A.~Affolder$^{51}$, 
Z.~Ajaltouni$^{5}$, 
J.~Albrecht$^{9}$, 
F.~Alessio$^{37}$, 
M.~Alexander$^{50}$, 
S.~Ali$^{40}$
% ...

{\footnotesize \it
% ...
$ ^{5}$Clermont Universit\'{e}, Universit\'{e} Blaise Pascal, CNRS/IN2P3, LPC, Clermont-Ferrand, France\\
$ ^{6}$CPPM, Aix-Marseille Universit\'{e}, CNRS/IN2P3, Marseille, France\\
% ...
$ ^{9}$Fakult\"{a}t Physik, Technische Universit\"{a}t Dortmund, Dortmund, Germany\\
$ ^{10}$Max-Planck-Institut f\"{u}r Kernphysik (MPIK), Heidelberg, Germany\\
% ...
$ ^{36}$Universidad de Santiago de Compostela, Santiago de Compostela, Spain\\
$ ^{37}$European Organization for Nuclear Research (CERN), Geneva, Switzerland\\
% ...
$ ^{40}$Nikhef National Institute for Subatomic Physics, Amsterdam, The Netherlands\\
% ...
$ ^{45}$H.H. Wills Physics Laboratory, University of Bristol, Bristol, United Kingdom\\
% ...
$ ^{50}$School of Physics and Astronomy, University of Glasgow, Glasgow, United Kingdom\\
$ ^{51}$Oliver Lodge Laboratory, University of Liverpool, Liverpool, United Kingdom\\
% ...
$ ^{53}$School of Physics and Astronomy, University of Manchester, Manchester, United Kingdom\\
% ...
}
\end{flushleft}


\cleardoublepage


\renewcommand{\thefootnote}{\arabic{footnote}}
\setcounter{footnote}{0}



\pagestyle{plain} 
\setcounter{page}{1}
\pagenumbering{arabic}

\linenumbers

%


\section{Introduction}
\label{sec:Introduction}

The \Bc~meson is of special interest, 
as it is the only meson consisting of two heavy quarks 
of different flavours.
It is the~heaviest meson
that decays through 
weak interactions, 
with either the~\cquark~or $\bar{\bquark}$~quark decaying
or through their weak~annihilation~\cite{Chang:1992pt,*Gershtein:1994jw,
*Gershtein:1997qy,*Colangelo:1999zn,*Kiselev:2003mp,Likhoded:2009ib,*Likhoded:2010jr}. 
Although the \Bc~meson was discovered in 1998 by the CDF 
collaboration~\cite{CDF_Bc,*Abe:1998fb}, 
relatively few decay channels were observed~\cite{Aaltonen:2007gv,*Abazov:2008kv}
prior to LHCb~measurements~\cite{LHCb-PAPER-2011-044,LHCb-PAPER-2012-054,
LHCb-PAPER-2013-010,
LHCb-PAPER-2013-044,
LHCb-PAPER-2013-021}.

In the factorisation approximation~\cite{Bauer:1986bm,*Wirbel:1988ft}, 
the~$\Bc\to\jpsi\Kp\Km\pip$~decay\footnote{The inclusion of 
charge conjugate modes is implicit throughout this paper.} 
is characterised  by the~form factors of the~$\Bc\to\jpsi\W^{+}$
transition and the~spectral functions
for the~subsequent hadronisation
of the~virtual $\W^{+}$~boson into light 
hadrons~\cite{Likhoded:2009ib,*Likhoded:2010jr}.
A~measurement of the~branching fractions of exclusive 
\Bc~meson decays into final states consisting 
of charmonium and light hadrons allows the validity of 
the~factorisation theorem to be tested.
Similar studies of factorisation have 
been performed on 
$\B\to\D^{\left(\ast\right)}\Km\kaon^{*0}$~decays~\cite{Drutskoy:2002ib}.
The~predictions for the~ratio of branching fractions 
$\BR\left(\Bc\to\jpsi\Kp\Km\pip\right)/\BR\left(\Bc\to\jpsi\pip\right)$
are 0.49 and 0.47~\cite{Lesha}, 
using form factor contributions from Refs.~\cite{Kis1}
and~\cite{Ebert}, respectively.

In this article, the~first observation of the~decay 
$\Bc\to\jpsi\Kp\Km\pip$~and 
a~measurement of 
$\BR(\Bc\to\jpsi\Kp\Km\pip)/\BR(\Bc\to\jpsi\pip)$~are 
reported. 
The~analysis is based on proton-proton ($\proton\proton$)~collision data, 
corresponding to an~integrated luminosity of~1\invfb
at a~centre-of-mass energy of 7\tev and 2\invfb at 8\tev,
collected with the LHCb detector.

\section{Detector and software}
\label{sec:Detector}

The \lhcb detector~\cite{Alves:2008zz} is a single-arm forward
spectrometer covering the \mbox{pseudorapidity} range $2<\eta <5$,
designed for the study of particles containing $\mathrm{b}$ or $\mathrm{c}$
quarks. The~detector includes a high-precision tracking system
consisting of a silicon-strip vertex detector surrounding the 
$\proton\proton$~interaction region, 
a~large-area silicon-strip detector located
upstream of a dipole magnet with a bending power of about
$4{\rm\,Tm}$, and three stations of silicon-strip detectors and straw
drift tubes placed downstream.
The combined tracking system provides a momentum measurement with
relative uncertainty that varies from 0.4\% at 5\gevc to 0.6\% at 100\gevc,
and impact parameter resolution of 20\mum for
tracks with high transverse momentum. Charged hadrons are identified
using two ring-imaging Cherenkov detectors~\cite{LHCb-DP-2012-003}. 
Muons are identified by a system composed of alternating layers 
of iron and multiwire proportional chambers~\cite{LHCb-DP-2012-002}. 
The trigger~\cite{LHCb-DP-2012-004} 
consists of a hardware stage, based on information from the calorimeter 
and muon systems, followed by a software stage, which applies a full event
reconstruction. 

This analysis uses events collected by triggers that select 
the~\mumu~pair from the~\jpsi~meson decay with high efficiency. 
At the hardware stage either one or two muon  candidates are required. 
In the case of single muon triggers, the transverse  momentum, $\pt$, 
of the candidate is required to be greater than $1.5\gevc$. 
For dimuon  candidates, the product of the \pt~of muon 
candidates is required to satisfy~$\sqrt{\pt_1\pt_2}>1.3\gevc$. 
At the subsequent software trigger stage, two muons with 
invariant mass in the interval $2.97<m_{\mumu}<3.21\gevcc$, and 
consistent with originating from a~common vertex, are required.

Simulated~$\proton\proton$~collisions are generated using 
\pythia~6.4~\cite{Sjostrand:2006za} with the 
configuration described in Ref.~\cite{LHCb-PROC-2010-056}.
Final-state QED radiative corrections are included using 
the \photos~package~\cite{Golonka:2005pn}.
The \Bc~mesons are produced by a dedicated generator, 
\bcvegpy~\cite{BCVEGPY}. The decays of all hadrons are performed 
by \evtgen~\cite{Lange:2001uf}, and a specific model is implemented 
to generate the decays of $\Bc\to\jpsi\Kp\Km\pip$, assuming 
factorisation~\cite{Lesha}. 
The model has different 
$\Bc\to\jpsi$~form factors implemented, 
calculated using QCD sum rules~\cite{Kis1} or
using a~relativistic quark model~\cite{Ebert}.
These model predictions are very similar 
and those based on the latter 
are used in the~simulation. 
The coupling of $\Kp\Km\pip$ to the~virtual \Wp~is taken 
from $\Ptau$~decays~\cite{Lee:2010tc}, 
following Refs.~\cite{Kuhn:1990ad,Likhoded:2009ib,Berezhnoy:2011is,*Berezhnoy:2011nx,*Luchinsky:2012rk,*Likhoded:2013iua},
and modelled through the intermediate
$\mathrm{a}^+_{1}\to \overline{\kaon}^{*0}\Kp(\overline{\kaon}^{*0}\to\Km\pip)$~decay chain.
The~interaction of the~generated particles with the~detector 
and its response are implemented using 
the~\geant~toolkit~\cite{Allison:2006ve, *Agostinelli:2002hh} as described in
Ref.~\cite{LHCb-PROC-2011-006}.

%

\section{Candidate selection}
\label{sec:EventSelection}

The signal $\Bc\to\jpsi\Kp\Km\pip$~and normalisation $\Bc\to\jpsi\pip$~decays  
are reconstructed using the $\jpsi\to\mumu$~channel. 
Common selection criteria are used in both channels
with additional requirements to identify kaon 
candidates in the signal channel.

Muons are selected by requiring that the difference in logarithms 
of the~muon hypothesis likelihood
with respect 
to the pion hypothesis likelihood, 
$\Delta\ln \mathcal{L}_{\Pmu/\Ppi}$~\cite{LHCb-DP-2013-001,LHCb-DP-2012-002},
is greater than zero.
To select kaons\,(pions) the~corresponding difference 
in the~logarithms of likelihoods of the~kaon and pion hypotheses~\cite{LHCb-DP-2012-003}
is required to satisfy \mbox{$\Delta\ln\mathcal{L}_{\kaon/\Ppi}>2~\,(<0)$}. 

To ensure that they do not originate  
from a~$\proton\proton$~interaction vertex (PV), 
hadrons  must have 
$\chisq_{\mathrm{IP}}>4$, where 
$\chisq_{\mathrm{IP}}$~is defined
as the difference in~\chisq of a given PV 
reconstructed with and without the considered hadron.
When more than one PV
is reconstructed, that with the smallest value 
of $\chisq_{\mathrm{IP}}$~is chosen. 

Oppositely-charged muons that have a~transverse momentum greater 
than $0.55\gevc$ and that originate from  a common vertex 
are paired to form \jpsi~candidates. 
The quality of the vertex is ensured by requiring 
that the~\chisq of the vertex fit ($\chisqvtx$) 
is less than 20. The vertex is required to be well-separated from the 
reconstructed PV by selecting candidates with decay length
significance greater than~3. 
The~invariant mass of the~\jpsi~candidate 
is required to be between 3.020 and 3.135$\gevcc$. 

The selected $\jpsi$~candidates are then combined 
with a~\pip~meson candidate or 
a~$\Kp\Km\pip$~combination to 
form \Bc~candidates. The quality of the common vertex 
is ensured by requiring $\chisqvtx<35\,(16)$ 
for the~signal\,(normalisation) channel, and 
that the~$\chisq$~values for the distance of 
closest approach for the~$\Kp\Km$, 
$\Km\pip$~and $\Kp\pip$~combinations are less than~9. 
To suppress the~combinatorial background, the~kaons\,(pions) 
are required to have \mbox{$\pt>0.8\,(0.5)\gevc$}.  
To improve the invariant mass resolution a~kinematic 
fit~\cite{Hulsbergen:2005pu} 
is performed. The~invariant mass of the~\jpsi~candidate 
is constrained to the~known value of \jpsi~mass~\cite{PDG2012},
the decay products of the~\Bc~candidate are required to originate 
from a~common vertex, and the~momentum vector of the~\Bc~candidate is 
required to point to the~PV.
When more than one PV
is reconstructed, that with the smallest value 
of $\chisq_{\mathrm{IP}}$~is chosen. 
The~$\chisq$ per degree of freedom 
for this fit is required to be less than~5.  
This requirement also reduces the potential contamination from 
decay chains with intermediate long-lived particles, namely 
$\Bc\to\jpsi\Dsp$,
$\Bc\to\Bs\pip$ and 
$\Bc\to\Bu\Km\pip$,
followed by 
$\Ds\to\Kp\Km\pip$,
$\Bs\to\jpsi\Kp\Km$ and 
$\Bu\to\jpsi\Kp$, respectively.
To reduce contributions from  
the known~$\Bc\to\jpsi\Ds$~\cite{LHCb-PAPER-2013-010} and 
$\Bc\to\Bs\pip$~decays~\cite{LHCb-PAPER-2013-044}
to a~negligible level,  the~invariant masses of the~$\Kp\Km\pip$~and 
$\jpsi\Kp\Km$~systems are required to differ 
from the~known \Ds~and \Bs~masses~\cite{LHCb-PAPER-2013-011,PDG2012}
by more than $18$~and $51\mevcc$, respectively,
corresponding  to~$\pm3\Psigma$, where $\Psigma$~is the mass 
resolution of the~intermediate state.
The decay time of the \Bc~candidate ($ct$)
is required to be between 
$150\mum$ and $1\mm$.
The upper limit corresponds to 
approximately 7~lifetimes of the~\Bc~meson.

%
%
\section{Signal and normalisation yields}
\label{sec:Nratio1}

\begin{figure}[t]
  \setlength{\unitlength}{1mm}
  \centering
   \begin{picture}(150,60)
    \put(0,0){
      \includegraphics*[width=75mm,height=60mm,%
      ]{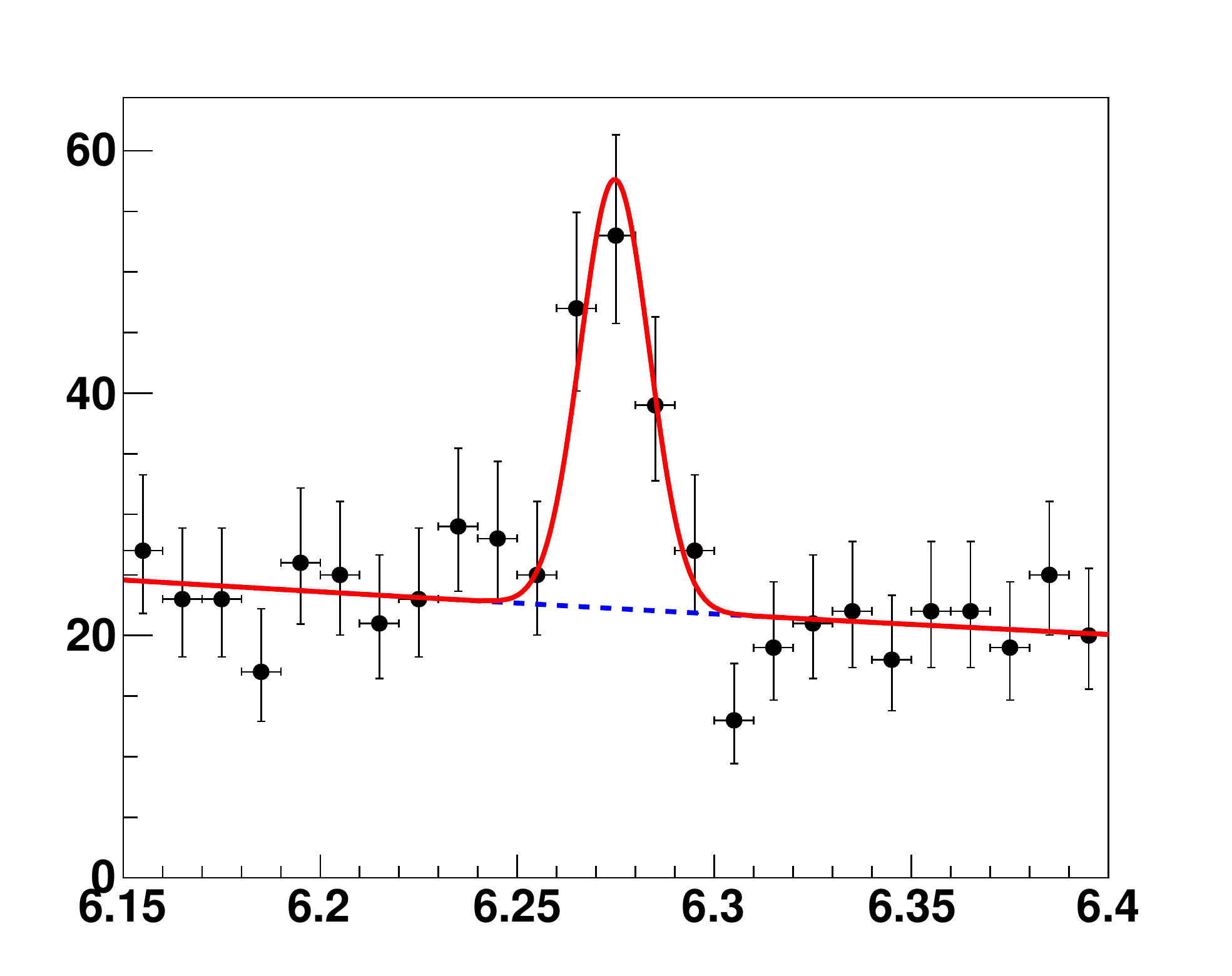}
    }
    \put(75,0){
      \includegraphics*[width=75mm,height=60mm,%
      ]{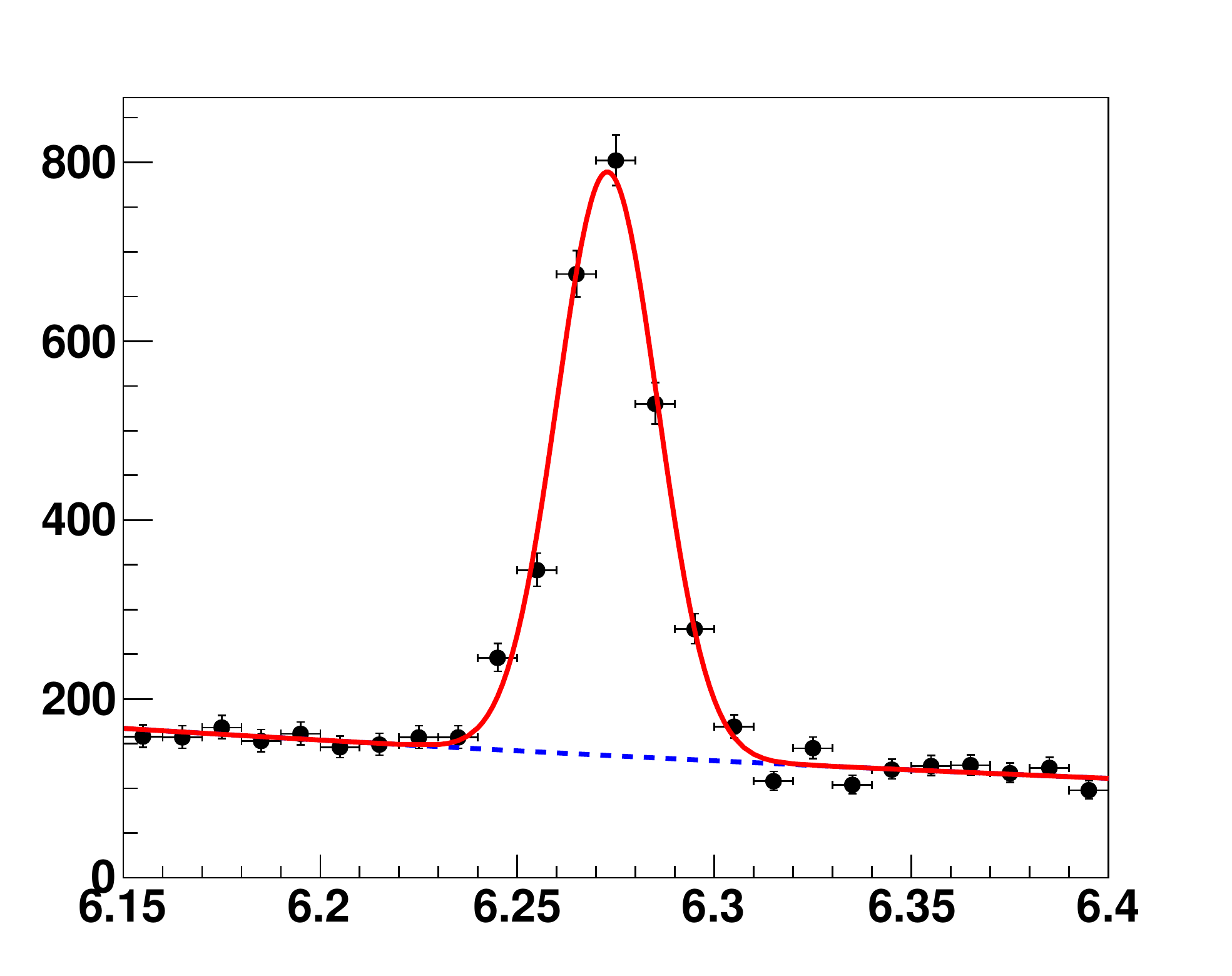}
    }
    \put(-2, 13 ) { \small
      \begin{sideways}%
        Candidates/(10\mevcc)
      \end{sideways}%
    }
    \put(72.5,13 ) { \small
      \begin{sideways}%
        Candidates/(10\mevcc)
      \end{sideways}%
    }
    \put(28,-1)   { $m_{\jpsi\Kp\Km\pip}$ }
    \put(52,-1)   { $\left[ \mathrm{GeV}/c^2\right]$}
    \put(107,-1)  { $m_{\jpsi\pip}$}
    \put(128,-1)  { $\left[ \mathrm{GeV}/c^2\right]$}
    \put(  50,45  ){ LHCb}
    \put( 125,45  ){ LHCb}
    \put(  15,45  ){ (a)}
    \put(  90,45  ){ (b)}
  \end{picture}
  \caption { \small
    Mass distribution for selected (a)~$\Bc\to\jpsi\Kp\Km\pip$
    and (b)~$\Bc\to\jpsi\pip$~candidates. 
    The result of the~fit described in the~text is
    superimposed (solid line) together with the background 
    component (dashed line).
  }
  \label{fig:Fig_1}
\end{figure}
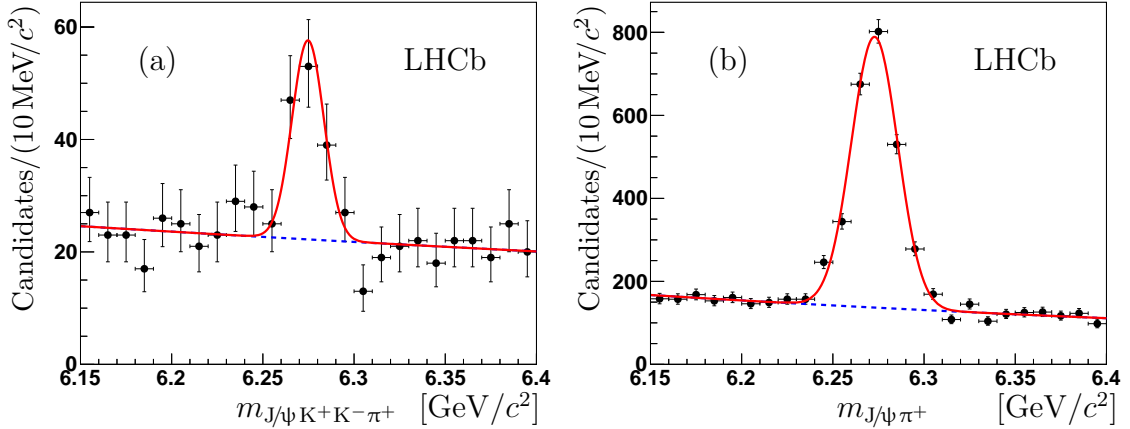

The invariant mass distribution of the selected 
$\Bc\to\jpsi\Kp\Km\pip$~candidates 
is shown in Fig.~\ref{fig:Fig_1}(a). 
To estimate the signal yield, $N_{\mathrm{S}}$, 
an extended unbinned maximum 
likelihood fit to the mass distribution is performed.
The \Bc~signal is modelled by a~Gaussian distribution 
and the~background
by an~exponential function. 
The values of the signal parameters obtained 
from the fit are summarised in Table~\ref{tab:signal_fitres_kkpi} and 
the result is shown in Fig.~\ref{fig:Fig_1}(a).
The~statistical significance of the observed signal yield 
is calculated as 
\mbox{$\sqrt{2\Delta\ln\mathcal{L}}$}, 
where $\Delta\ln\mathcal{L}$~is the~change in the~logarithm of 
the~likelihood function when the~signal component
is excluded from the~fit, 
relative to the~default fit,
and is found to be 6.3~standard deviations. 

\begin{table}[b]
\centering
\caption{\small
  Parameters of the signal function of the  
  fit to the $\jpsi\Kp\Km\pip$~mass distribution. 
  Uncertainties are statistical only.}
\vspace*{3mm}
\begin{tabular*}{0.50\textwidth}{@{\hspace{5mm}}lc@{\extracolsep{\fill}}c@{\hspace{5mm}}}
  \multicolumn{2}{c}{~~Parameter}  & Value \\ \hline
  $m_{\Bc} $                        &  $\left[\mevcc\right]$ & $6274.8 \pm 1.7\phantom{000}$   \\
  $\Psigma_{\Bc}                   $ &  $\left[\mevcc\right]$ & $\phantom{0}8.8     \pm 1.5\phantom{0}$     \\
  $N_{\mathrm{S}}                    $ &                        & $78    \pm 14$     \\
\end{tabular*}
\label{tab:signal_fitres_kkpi}
\end{table}

The invariant mass distribution of the selected 
$\Bc\to\jpsi\pip$~candidates is shown in~Fig.~\ref{fig:Fig_1}(b). 
To estimate the signal yield, an~extended unbinned maximum likelihood fit to the 
mass distribution is performed, where the \Bc~signal is modelled by 
a~Gaussian distribution and the background by an exponential function. 
The fit gives a yield of $2099 \pm 59$~events. 

\begin{figure}[!t]
  \setlength{\unitlength}{1mm}
  \centering
  \begin{picture}(150,180)
    \put(0,120){
      \includegraphics*[width=75mm,height=60mm,%
      ]{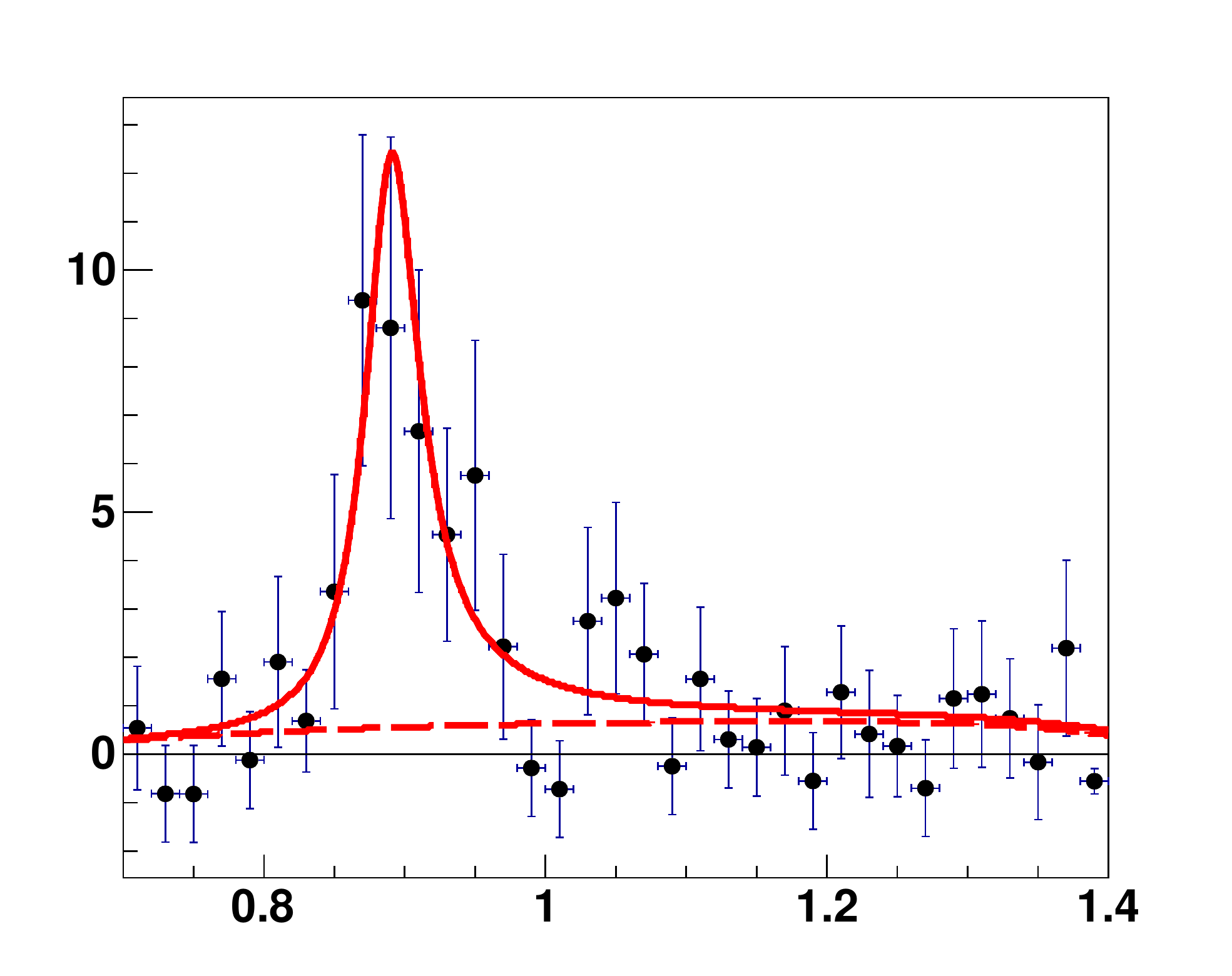}
    }
    \put(77,120){
      \includegraphics*[width=75mm,height=60mm,%
      ]{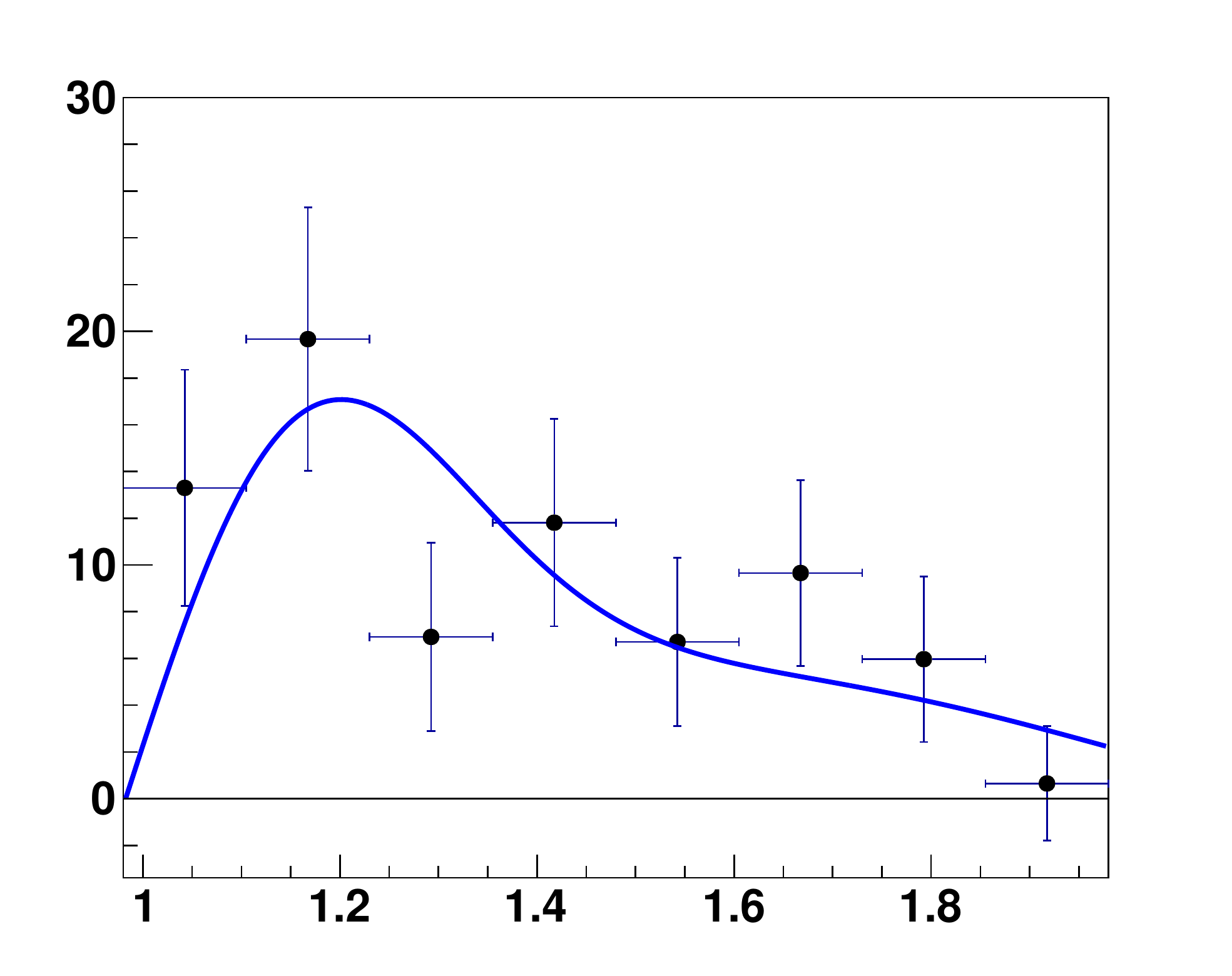}
    }
    \put(0,60){
      \includegraphics*[width=75mm,height=60mm,%
      ]{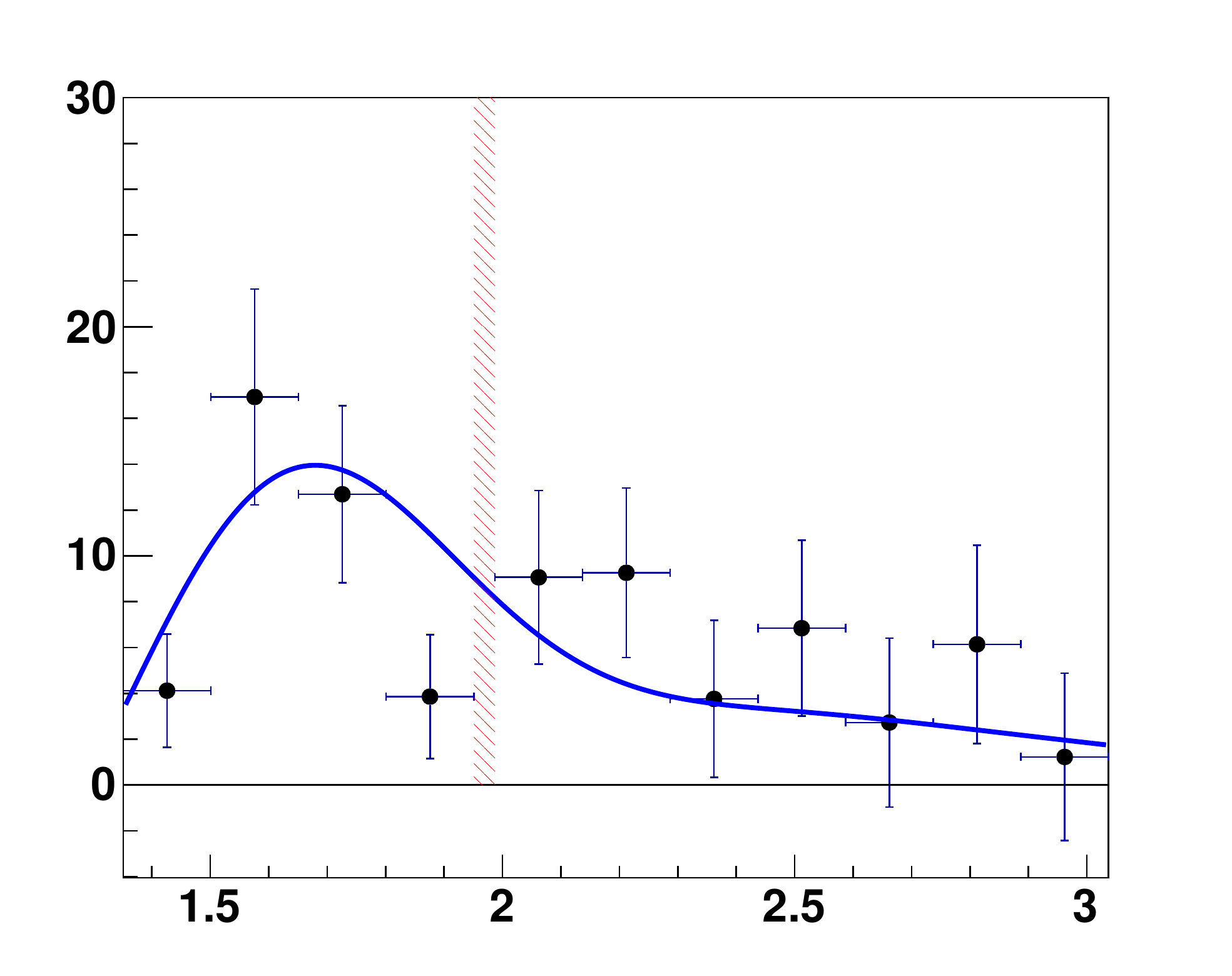}
    }
    \put(77,60){
      \includegraphics*[width=75mm,height=60mm,%
      ]{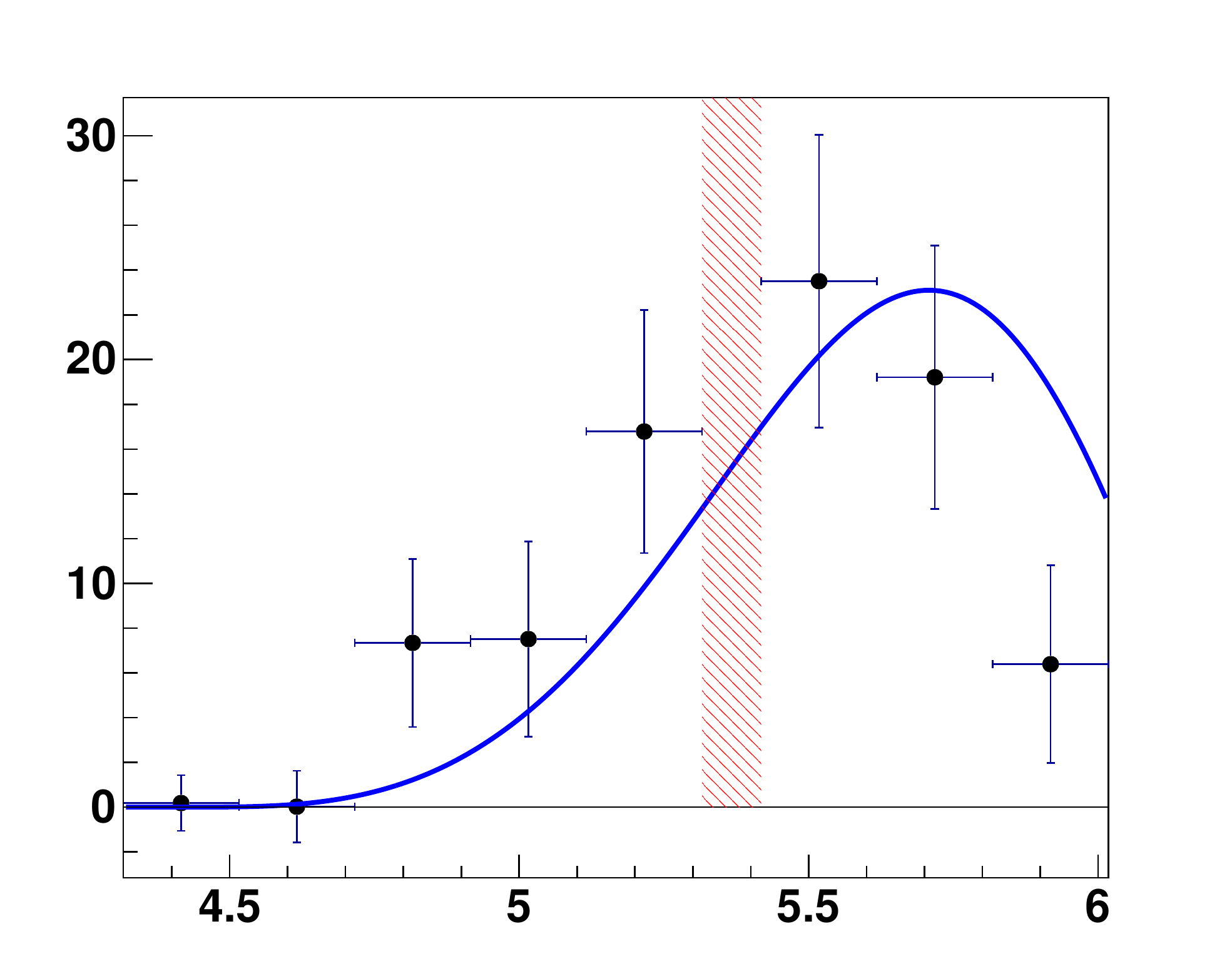}
    }
    \put(0,0){
      \includegraphics*[width=75mm,height=60mm,%
      ]{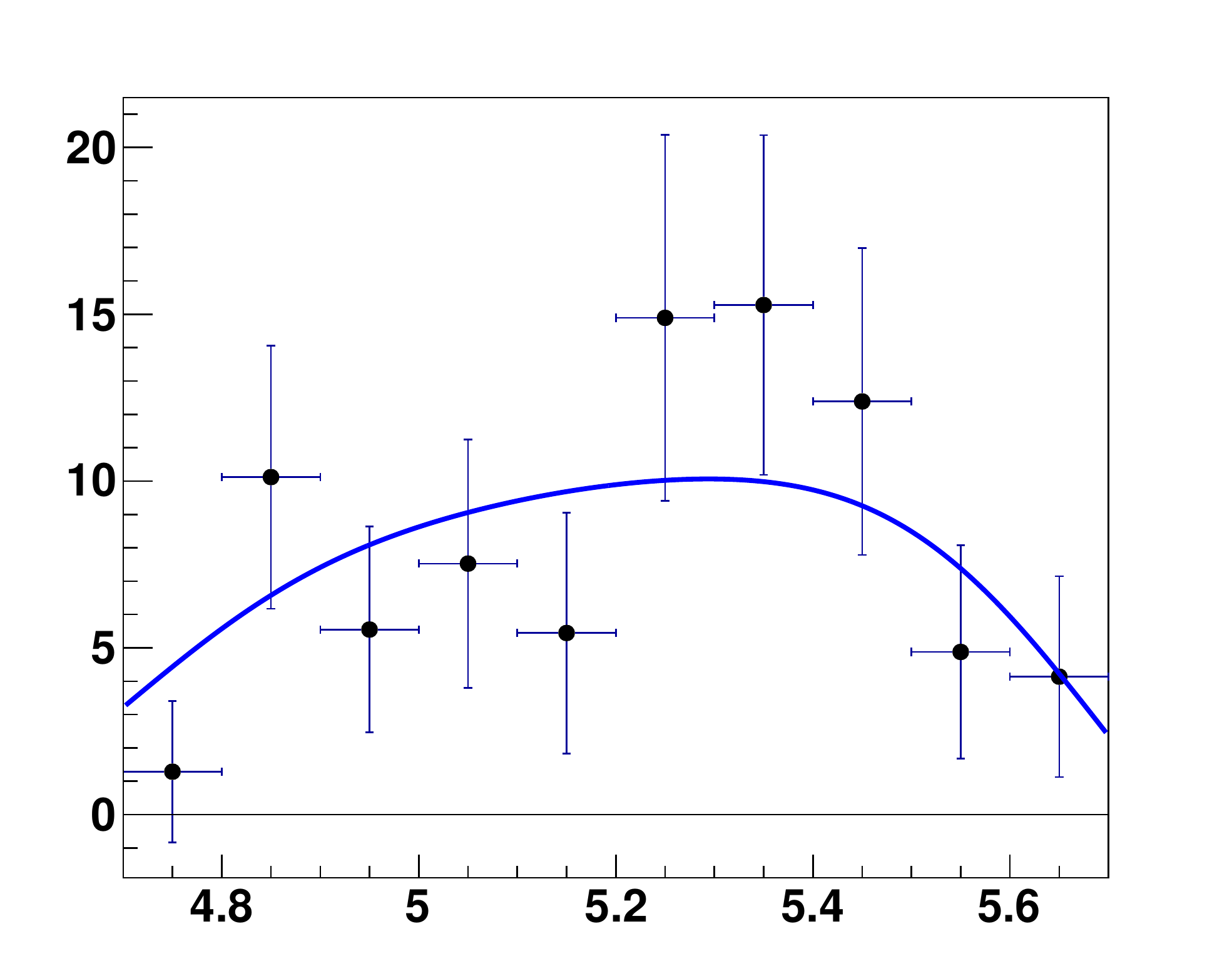}
    }
    \put(77,0){
      \includegraphics*[width=75mm,height=60mm,%
    ]{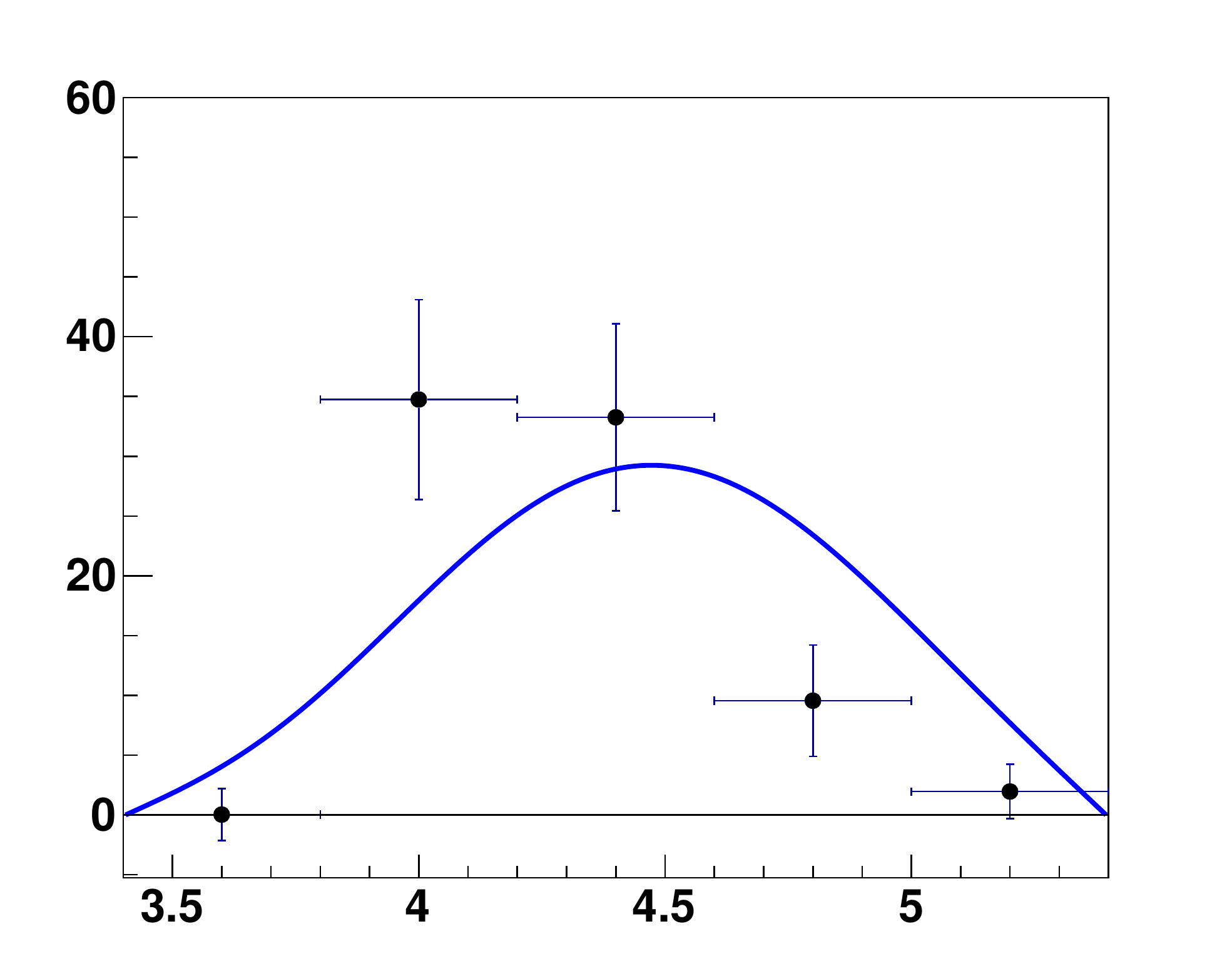}
    }
    \put(30,119)  { $m_{\Km\pip}$ }
    \put(52,119)  { $\left[ \mathrm{GeV}/c^2\right]$}
    \put(107,119)  { $m_{\Kp\Km}$}
    \put(129,119)  { $\left[ \mathrm{GeV}/c^2\right]$}
    \put(30,-1)   { $m_{\jpsi\Km\pip}$}
    \put(52,-1)   { $\left[ \mathrm{GeV}/c^2\right]$}
    \put(30,59)   { $m_{\Kp\Km\pip}$ }
    \put(52,59)   { $\left[ \mathrm{GeV}/c^2\right]$}
    \put(107,-1)  { $m_{\jpsi\Kp}$}
    \put(129,-1)  { $\left[ \mathrm{GeV}/c^2\right]$}
    \put(107,59)  { $m_{\jpsi\Kp\Km}$}
    \put(129,59)  { $\left[ \mathrm{GeV}/c^2\right]$}
    \put(-1 , 130 ) { \small
      \begin{sideways}%
        Candidates/(20\mevcc)
      \end{sideways}%
    }
    \put(76 , 130 ) { \small
      \begin{sideways}%
       Candidates/(125\mevcc)
      \end{sideways}%
    }
    \put(-1 , 70 ) { \small
      \begin{sideways}%
        Candidates/(150\mevcc)
      \end{sideways}%
    }
    \put(-1 , 10 ) { \small
      \begin{sideways}%
       Candidates/(100\mevcc)
      \end{sideways}%
    }
    \put(76 , 70 ) { \small
      \begin{sideways}%
       Candidates/(200\mevcc)
      \end{sideways}%
    }
    \put(76 , 10 ) { \small
      \begin{sideways}%
       Candidates/(400\mevcc)
      \end{sideways}%
    }
    \put(128,167){ LHCb}
    \put(53,167) { LHCb} 
    \put(128,107){ LHCb}
    \put(53,107) { LHCb}
    \put(128,45) { LHCb}
    \put(53 ,45) { LHCb}
    \put(10,167 ){ (a) }
    \put(87,167 ){ (b) }
    \put(10,107 ){ (c) }
    \put(87,107 ){ (d) }
    \put(10,47  ){ (e) }
    \put(87,47  ){ (f) }
  \end{picture}
  \caption { \small
    Background-subtracted invariant mass distributions for 
   (a)~$\Km\pip$, 
   (b)~$\Kp\Km$, 
   (c)~$\Kp\Km\pip$, 
   (d)~$\jpsi\Kp\Km$, 
   (e)~$\jpsi\Km\pip$ and 
   (f)~$\jpsi\Kp$ in $\Bc\to\jpsi\Kp\Km\pip$~decay. 
   The (red) full line in the $\Km\pip$~mass distribution (a) is composed
   of a~resonant $\overline{\kaon}^{*0}$~contribution and 
   a~non-resonant component indicated by the dashed line.
   The (blue) full line in (b)--(f) shows 
   the predictions of the model~\cite{Lesha} used in the~simulation.
   The~regions $\pm 18\mevcc$ around the~$\Dsp$~mass and 
   $\pm 51\mevcc$ around the~$\Bs$~mass are 
   excluded from the~analysis and are indicated 
   by the shaded areas on (c) and (d), respectively.  
}
  \label{fig:splot}
\end{figure}
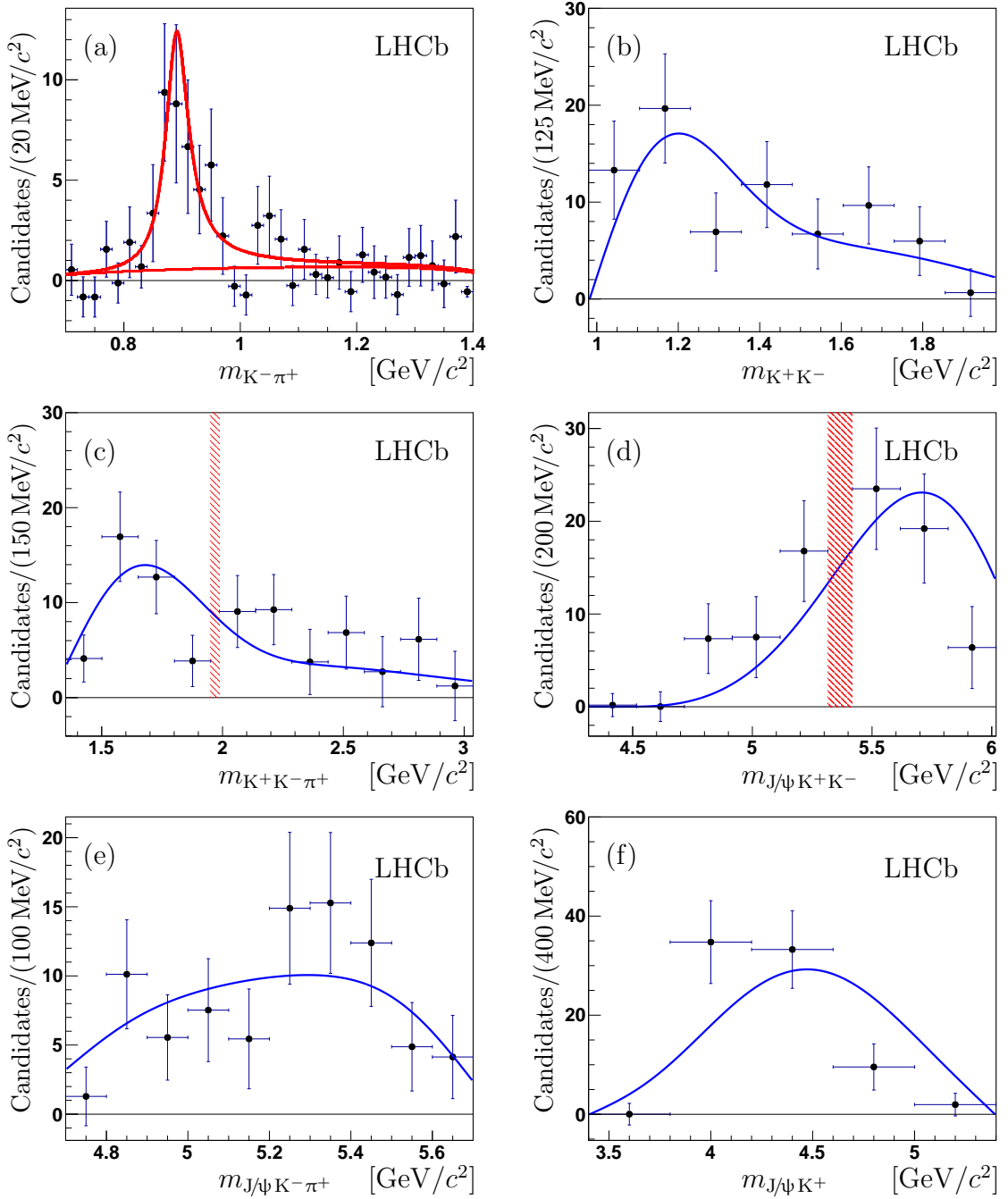

For $\Bc\to\jpsi\Kp\Km\pip$~candidates, 
the resonant structures in
the~$\Km\pip$, $\Kp\Km$, $\Kp\Km\pip$, 
$\jpsi\Kp\Km$, $\jpsi\Km\pip$ and $\jpsi\Kp$~systems are studied and 
the~possible contributions from 
the~decays $\Bc\to\Bd\Kp$  and $\Bc\to\Bu\Km\pip$, followed by 
subsequent decays $\Bd\to\jpsi\Km\pip$ and $\Bu\to\jpsi\Kp$~are investigated. 
 The~\sPlot~technique~\cite{Pivk:2004ty} 
is used to subtract 
the estimated background contribution from 
the~corresponding mass distributions. 
The~results are shown in Fig.~\ref{fig:splot}. 

The binned $\Km\pip$~invariant mass distribution, presented in Fig.~\ref{fig:splot}(a), 
is fitted with the~sum of two components, one representing the~$\overline{\mathrm{K}}^{*0}$~resonance 
and a~non-resonant component modelled with the LASS parametrisation~\cite{Aston:1987ir}. 
The~resonant component is described by a~relativistic 
P-wave Breit-Wigner function. The form factor
for the~$\left(1^-\right)\to\left(0^-\right)\left(0^-\right)$~decay 
is taken from lowest order perturbation theory~\cite{Jackson:1964zd}, 
while the peak position and the natural width are fixed to their 
known values~\cite{PDG2012}. 
The~resulting resonant yield is $44 \pm 10$~decays,
where the uncertainty is statistical only.

Figures~\ref{fig:splot}(b)--(f) show the invariant 
mass distributions for 
the $\Kp\Km$, $\Kp\Km\pip$, 
$\jpsi\Kp\Km$, $\jpsi\Km\pip$ and 
$\jpsi\Kp$~final states. In~contrast
to Fig.~\ref{fig:splot}(a),
no narrow structures are visible.
The~predictions from the~model of Ref.~\cite{Lesha}
are also presented in Fig.~\ref{fig:splot},
and are found to give an~acceptable 
description of the~data. 

\afterpage{\clearpage}

%

\section{Efficiency and systematic uncertainties}
\label{sec:syst}
As the ratio of branching fractions is measured, 
many  potential sources of systematic uncertainty cancel
in the ratio of efficiencies for the normalisation and signal decays.
The overall efficiency for both decays is the product of 
the~geometrical acceptance of the~detector, 
reconstruction, selection and trigger
efficiencies. These are estimated using simulation
and the~ratio of the efficiencies is found to be  
\begin{equation*}
  \dfrac{\Pvarepsilon(\Bc\to\jpsi\pip)}{\Pvarepsilon(\Bc\to\jpsi\Kp\Km\pip)} = 14.3 \pm 0.4,
\end{equation*}
where the uncertaintty is statistical only.
Systematic uncertainties that do not cancel 
in this ratio are discussed below
and summarised in Table~\ref{table:syst}.


The main uncertainty arises from the imperfect knowledge 
of the shape of the signal and background components used to model the 
$\Bc$~mass distributions. It is estimated using 
an~alternative model 
to describe the~$\Bc\to\jpsi\Kp\Km\pip$~and $\Bc\to\jpsi\pip$~mass 
distributions consisting of a~Crystal Ball function~\cite{Skwarnicki:1986xj} 
for the signal and a~linear function for the background. 
The~changes in the yields relative to the default fits are used 
to determine a~5.0\,\%~uncertainty on the~number of signal 
candidates in both channels, and is dominated by the~large 
background level in signal decay.


Other systematic uncertainties arise from differences
between data and simulation in the~track reconstruction 
efficiency for charged particles. 
The largest of these arises from the knowledge of the hadronic 
interaction probability in the detector, which has an~uncertainty 
of~$2.0\,\%$ per track~\cite{LHCb-PAPER-2010-001}. 
Further uncertainties related to the recontruction of 
charged kaons contribute 
0.6\,\%~per kaon~\cite{LHCb-PAPER-2013-010,LHCb-PUB-2011-025,*LHCb-DP-2013-002}. 
%
%
The differences in the kinematic properties 
of the charged pion in the signal and 
normalisation channels  are also considered 
as a~source of systematic uncertainty.
The total uncertainty assigned to track reconstruction 
and selection is~$4.2\,\%$.


The systematic uncertainty associated with kaon identification 
is studied using a~kinematically similar sample of reconstructed 
$\Bu\to\jpsi\left(\Kp\Km\right)_{\Pphi}\Kp$~decays~\cite{LHCb-PAPER-2013-010}. 
An~uncertainty of $3.0\,\%$ is assigned.


A source of systematic uncertainty arises from the potential 
disagreement between data and simulation in the efficiencies 
of the selection criteria. 
To study this  effect, 
the criteria are varied to values that correspond to a~$20\,\%$~change 
in the~signal yields.
The~variation of the~relative difference between 
data and simulation on the~number of selected signal candidates 
reaches~1.6\,\%, which is assigned as a~systematic uncertainty
from this source, and includes effects related to 
pion identification criteria.


The dependence of the $\Bc\to\jpsi\Kp\Km\pip$~decay reconstruction 
and selection efficiency on the decay model 
implemented in the simulation 
is estimated from a comparison of the~$\Kp\Km\pip$~invariant 
mass distributions in data and simulation,
which has the greatest dependence on the decay model. 
This combined efficiency is recomputed 
after reweighting the~$\Kp\Km\pip$~mass distribution to that 
observed in data. The~relative difference of $2.5\,\%$ 
observed is taken as the~systematic uncertainty due to 
the~decay model.


Other systematic uncertainties are related to the widths 
of the $\Kp\Km\pip$~and $\jpsi\Kp\Km$~mass regions vetoed 
in the analysis to reject contributions from $\Bc\to\jpsi\Ds$~and 
$\Bc\to\Bs\pip$~decays. 
These are estimated by varying the widths of the vetoed regions 
and recomputing the $\Bc\to\jpsi\Kp\Km\pip$~signal yields, 
taking into account the changes in efficiency. 
A~systematic uncertainty of $1.0\,\%$ is assigned.  


The efficiency of the requirement on the \Bc~decay time depends on
the~value of the~\Bc~lifetime used in the~simulation. 
The~decay time distributions for simulated events 
are reweighted after changing the 
$\Bc$~lifetime by one standard deviation around 
the known value~\cite{PDG2012}, as well as using the lifetime value
recently measured by the CDF collaboration~\cite{CDFtime}, 
and the efficiencies are recomputed. 
The observed 2.5\,\%~variation in the~ratio 
of efficiencies is used as the~systematic uncertainty.


The agreement of the absolute trigger efficiency between data and simulation 
has been validated to a precision of 4\,\%~using the technique described 
in Refs.~\cite{LHCb-DP-2012-004,LHCb-PAPER-2011-013,LHCb-PAPER-2010-001} 
with a~large sample of 
$\Bp\to\jpsi\left(\Kp\Km\right)_{\Pphi}\Kp$~events~\cite{LHCb-PAPER-2013-010}.  
A further cancellation of uncertainties 
in the ratio of branching fractions has been tested with the~high statistics decay 
modes $\Bp\to\jpsi\Kp$~and $\Bp\to\Ppsi(2\PS)\Kp$~\cite{LHCb-PAPER-2012-010}, 
resulting in a systematic uncertainty of 1.1\,\%.


Potential uncertainties related to the stability of the data taking 
conditions are tested by studying the 
ratio of the yields of $\Bu\to\jpsi\Kp\pip\pim$~and 
$\Bu\to\jpsi\Kp$~decays for 
different data taking periods. 
According to this study an additional systematic 
uncertainty of $2.5\,\%$ is assigned~\cite{LHCb-PAPER-2013-010}.     
The final source  of systematic uncertainty considered 
originates from the~dependence of the geometrical acceptance
on both the~beam crossing angle and the~position of 
the~luminous region. 
The~observed difference in the efficiency ratios is 
taken as an estimate of the systematic 
uncertainty and is~$0.4\,\%$.
The correlation between  this uncertainty and  
the~previous one  is neglected.

\begin{table}[tb]
  \centering
  \caption{ \small
    Relative systematic uncertainties
    for the ratio of branching fractions
    of \mbox{$\Bc\to\jpsi\Kp\Km\pip$} and \mbox{$\Bc\to\jpsi\pip$}.
    The total uncertainty is the~quadratic sum 
    of the individual components. 
  } \label{table:syst}
  \vspace*{3mm}
  \begin{tabular*}{0.8\textwidth}{@{\hspace{10mm}}l@{\extracolsep{\fill}}c@{\hspace{10mm}}}
    Source & Uncertainty~$\left[\%\right]$
    \\
    \hline
    Fit model                           &  5.0
    \\
    Track reconstruction and selection	&  4.2
    \\
    Kaon  identification                &  3.0
    \\
    Data and simulation disagreement    &  1.6
    \\
    Decay model dependence   &  2.5
    \\
    Vetoed mass intervals & 1.0
    \\
    $\Bc$ lifetime & 2.5
    \\
    Trigger                           &  1.1
    \\
    Stability of data taking conditions & 2.5
    \\
    Geometrical acceptance            &  0.4
    \\
    \hline
    Total    &  8.7
  \end{tabular*}
 \end{table}



\section{Results and summary }
\label{sec:result}

The decay $\Bc\to\jpsi\Kp\Km\pip$ is observed for the first time,
and a~signal yield of $78 \pm 14$ is reported.
This analysis uses a~data sample corresponding to  an~integrated luminosity 
of~1\invfb
at a~centre-of-mass energy of 7\tev and 2\invfb at 8\tev.
The~significance, taking into account the systematic uncertainties 
due to the~fit function, peak position and mass resolution 
in the~default fit, is estimated to be 6.2~standard deviations.

Using the $\Bc\to\jpsi\pip$ mode as a~normalisation channel, 
the ratio of branching fractions is calculated as
\begin{equation*}
\dfrac{\BR\left(\Bc\to\jpsi\Kp\Km\pip\right)}
     {\BR\left(\Bc\to\jpsi\pip\right)} = 
\dfrac{N\left(\Bc\to\jpsi\Kp\Km\pip\right)}
      {N\left(\Bc\to\jpsi\pip\right)} 
\times 
\dfrac{\Pvarepsilon(\Bc\to\jpsi\pip)}{\Pvarepsilon(\Bc\to\jpsi\Kp\Km\pip)},
\end{equation*}
where $N$~is the number of reconstructed decays
obtained from the fit described in Sect.~\ref{sec:Nratio1}. 
The ratio of branching fractions is measured to be 
\begin{equation*}
\dfrac{\BR\left(\Bc\to\jpsi\Kp\Km\pip\right)}
      {\BR\left(\Bc\to\jpsi\pip\right)} =0.53\pm0.10\pm0.05,
\end{equation*}
where the first uncertainty is statistical and the second systematic. 
The largest contribution to the 
$\Bc\to\jpsi\Kp\Km\pip$~decay is found to be from 
$\Bc\to\jpsi\overline{\mathrm{K}}^{*0}\Kp$~decays. 
The~theoretical predictions for the~branching fraction ratio
of 0.49 and 0.47~\cite{Lesha}, using form factors 
from Refs.~\cite{Kis1} and~\cite{Ebert},
respectively, are found to be in good agreement with this measurement.

\section*{Acknowledgements}

\noindent 
We thank 
A.K.~Likhoded and 
A.V.~Luchinsky 
for fruitful discussions about the dynamics of \Bc~decays.
We express our gratitude to our colleagues in the CERN
accelerator departments for the excellent performance of the LHC. We
thank the technical and administrative staff at the LHCb
institutes. We acknowledge support from CERN and from the national
agencies: CAPES, CNPq, FAPERJ and FINEP (Brazil); NSFC (China);
CNRS/IN2P3 and Region Auvergne (France); BMBF, DFG, HGF and MPG
(Germany); SFI (Ireland); INFN (Italy); FOM and NWO (The Netherlands);
SCSR (Poland); MEN/IFA (Romania); MinES, Rosatom, RFBR and NRC
``Kurchatov Institute'' (Russia); MinECo, XuntaGal and GENCAT (Spain);
SNSF and SER (Switzerland); NAS Ukraine (Ukraine); STFC (United
Kingdom); NSF (USA). We also acknowledge the support received from the
ERC under FP7. The Tier1 computing centres are supported by IN2P3
(France), KIT and BMBF (Germany), INFN (Italy), NWO and SURF (The
Netherlands), PIC (Spain), GridPP (United Kingdom). We are thankful
for the computing resources put at our disposal by Yandex LLC
(Russia), as well as to the communities behind the multiple open
source software packages that we depend on.



\addcontentsline{toc}{section}{References}
\setboolean{inbibliography}{true}
\bibliographystyle{LHCb}
\bibliography{main,LHCb-PAPER,LHCb-CONF,LHCb-DP,local}

\end{document}